# Radiation-induced mobility of small defect clusters in covalent materials

Hao Jiang[1], Li He[2], Dane Morgan[1,2], Paul M. Voyles[1,2], Izabela Szlufarska[1,2,*]

[1]*Materials Science Program,* [2]*Department of Materials Science & Engineering,*

*University of Wisconsin, Madison, WI 53706*

*Abstract*

Although defect clusters are detrimental to electronic and mechanical properties of semiconductor materials, annihilation of such clusters is limited by their lack of thermal mobility due to high migration barriers. Here, we find that small clusters in bulk SiC (a covalent material of importance for both electronic and nuclear applications) can become mobile at room temperature under the influence of electron radiation. So far, direct observation of radiation-induced diffusion of defect clusters in bulk materials has not been demonstrated yet. This finding was made possible by low angle annular dark field (LAADF) scanning transmission electron microscopy (STEM) combined with non-rigid registration technique to remove sample instability, which enables atomic resolution imaging of small migrating defect clusters. We show that the underlying mechanism of this athermal diffusion is ballistic collision between incoming electrons and cluster atoms. Our findings suggest that defect clusters may be mobile under certain irradiation conditions, changing current understanding of cluster annealing process in irradiated covalent materials.

*Main text*

Defect clusters can form in covalent materials during ion implantation (e.g., in semiconductor applications) [1,2] or during irradiation (e.g., in nuclear reactor applications) [3-6]. Accumulation of such clusters is highly detrimental to electronic, optical, and mechanical properties of these materials [1,4,7]. While point defects can often be eradicated by mutual recombination or diffusion to defect sinks, clusters of defects in covalent materials are known to be immobile and resist annealing because of their high energy barriers to migration [8,9]. For SiC specifically, accelerated molecular dynamics (MD) simulations have shown that the barrier to migration of clusters as small as just three C interstitials is ~4.3 eV [8], which means that these clusters are immobile below 1,200 K on typical experimental annealing time scales (i.e., 1 hour whereas the cluster performs less than 1 hop/day). Such high barriers explain why models of defect evolution in SiC have assumed clusters to be immobile even at elevated temperatures (e.g., Ref. [10,11]) and why the nature of such clusters, their resistance to high-temperature annealing, and their effects on opto-electronic properties have been a subject of many discussions in the semiconductor literature [12,13].

In this letter, we report a direct observation of interstitial clusters diffusion in bulk SiC under the influence of electron radiation at room temperature. Although it is known that in metals radiation can lead to the one-dimensional diffusion of interstitial loops along the glide direction [14-16], the underlying mechanism is entirely different from the one discussed here. Unlike in ceramics, in metals interstitial loops have inherently relatively low migration barriers (on the order of 0.02 eV [17]) and therefore they are mobile at room temperature. These otherwise mobile loops can become trapped by pinning points, such as solute atoms, and the role of radiation is to release the loops from traps [16]. Thus in this case radiation is not the driving force for defect migration, but rather it allows interstitial loops to undergo thermal diffusion that would occur in the absence of trapping solutes. The situation is qualitatively different when it comes to ceramics where it is generally accepted that clusters of self-interstitials or vacancies are immobile at room temperature due to their high inherent migration barriers (typically between 4 and 7 eV [8,9,18]). Therefore, the observed diffusion of clusters in SiC is an athermal process induced by radiation, rather than triggered by radiation.

In our experiments, 4H-SiC sample was first irradiated by 1 MeV Kr at 600 °C at a flux of $2.5\times10^{12}$ atoms/(cm$^2$s) to a dose of $3\times10^{14}$ Kr atoms/cm$^2$, 4° off the [0001] direction, producing a peak damage of 0.4 displacement per atom (dpa) at 0.3 μm depth, as estimated using the Stopping and Range of Ions in Matter software [19]. STEM samples were prepared by wedge polishing and argon ion milling with beam energy of 3.5 keV first, then 2 keV, and finally 0.5 keV. Subsequently electron irradiation was conducted at the depth corresponding to 0.26 dpa at room temperature under 200 kV with a FEI Titan S-Twin, and under 60 kV with a FEI Titan G2 60-300, both of aberration corrected STEM. The electron flux for 200 keV irradiation is in the range of $2.58\times10^5 - 4.05\times10^6$ e$^-$/(nm$^2$s), and for 60 keV irradiation is $2.22\times10^5 - 1.99\times10^7$ e$^-$/(nm$^2$s). For a pixel size $r$ much smaller than the probe size, the electron dose per frame is $d = ct/r^2$ for pixel dwell time $t$, where $c$ is the beam current in e$^-$/s. We adjusted the flux primarily by changing the STEM pixel size and the probe current. Low angle annular dark field (LAADF) image series, were taken with 200 keV beam of semi-convergence angle 17.5 mrad, collection angle 23.0 - 115 mrad, or 60 keV beam of semi-convergence angle 25.1 mrad, collection angle 29.0 - 145 mrad, in 0.95 s per frame for 128 frames. Sample drift and instrumental instabilities were removed from the image series using non-rigid registration [20]. The majority of results reported here correspond to 200 keV, but 60 keV experiments were also conducted to test our hypotheses regarding the mechanisms underlying radiation-induced diffusion.

Figure 1(a) shows a low-angle annular dark-field (LAADF) (detector covering 17.5 to 34 mrad) STEM image of Kr-irradiated 4H-SiC. The irregular dark blobs, each of which covers a few atomic columns, are irradiation-induced defect clusters, visible due to their strain fields, which are emphasized in LAADF imaging [21]. Fig. 1(b) is a high-angle ADF (HAADF) (detector covering 54 to 270 mrad) image of the same region of the same sample. The lack of contrast confirms that the contrast ascribed to defect clusters does not arise from other sources such as surface roughness, oxidation, or hydrocarbon contamination. The average cluster diameter was found to be 0.85 ± 0.01 nm (corresponding to no more than 15 point defects [22]) and the cluster density was found to be $(9.3 \pm 0.8) \times 10^{23}$ m$^{-3}$. These clusters are believed to be of interstitial-type because of the lack of vacancy mobility in this temperature regime [11].

In the experiments performed under 200 keV electron irradiation, many of the defect clusters are found to be mobile. One example is illustrated in Fig. 2, where a defect cluster is shown to move over a series of LAADF STEM images (see also video S1 in Ref. [22]). The images in the series were aligned using our recently developed non-rigid registration technique [20], which separates instrumental effects (drift and instability) from the actual motion of defects. Averaging of similarly aligned series has demonstrated sub-pm precision in locating atomic positions [20]. Here, it renders the underlying crystal lattice motionless, making the motion of the defect clusters visible to the human eye and amenable to quantitative analysis. Three more examples of aligned STEM image series are available in videos S2-S4 in Ref. [22].

In Fig. 2(a), the cluster has moved from the position marked with a red circle to one marked by a white circle within the time of 94 s. For these mobile defect clusters, we track the position of the center of the clusters in non-rigid registration aligned STEM images frame by frame [22], and the trajectory of the cluster in Fig. 2(a) is shown in Figs. 2(b)-2(f). We then calculate the squared displacement $\Delta r^2$ of each cluster from its starting position as a function of tracking time $t$. $\Delta r^2$ is subsequently used to determine the diffusion coefficient $D$

$$D = \frac{\Delta r(t)^2}{2dt}, \qquad (1)$$

where $d$ is the dimensionality of the motion (here $d = 3$). Fig. 3(a) shows $\Delta r^2$ vs $t$ of the mobile defects illustrated in Fig. 2; in Fig. 3(a) the slope of the fitted line is $2d \times D$. In Fig. 3(b), we report averages of diffusion coefficients determined for different fluxes of 200 keV electrons from the trajectories of multiple defects. In general diffusivity increases with increasing electron flux. Detailed information on trajectories and diffusion coefficients for each mobile defect under 200 keV is reported in Table S1 in Ref. [22]. In contrast to the 200 keV experiments, within the experimental time scale (~10$^2$ s), we did not observe significant displacements of clusters under 60 keV electron radiation even at a high flux of $1.99 \times 10^7$ e$^-$/(nm$^2$s). This does not imply the diffusion coefficient of clusters under 60 keV radiations is necessarily zero, instead, it suggests the diffusion coefficient is too small to identify observable displacements within the experimentally accessible time scale. By assuming the un-observable displacement is less than the radius of the average cluster size (~0.42 nm), the un-observable diffusion coefficient range within the experimental time scale is calculated as $(0-3) \times 10^{-4}$ nm$^2$/s by using equation (1). This range is significantly lower than diffusion coefficients of clusters induced by 200 keV electron irradiation.

What is the mechanism responsible for the radiation-induced diffusion of clusters in SiC? One possibility, which has been often invoked to explain effects of radiation on accelerating point defect diffusion in ceramics, is that ionization of defects can lower their migration barriers. This phenomenon has been reported for instance in MgAl$_2$O$_4$ [23] and Al$_2$O$_3$ [23,24]. Ionization was also invoked to explain local bond-switching and recovery of a damaged zone (without diffusion and mass transport) in SiC [25,26]. However, ionization cannot explain our results: the ionization cross section for 60 keV electrons is higher than for 200 keV electrons [27], but the diffusivity is substantially lower. Here, we propose that radiation-induced diffusion of otherwise immobile interstitial clusters in SiC is a result of ballistic collision of incoming electrons with cluster atoms. During such collision, kinetic energy is transferred from the high-energy electron beam to interstitial atoms, assisting them in overcoming the energy barrier to migration athermally. A similar mechanism had been previously proposed for radiation effects on vacancy diffusion in lead [28] and more recently also for defect transformation and migration on surfaces[29] and in mono-layer graphene [30-33]. However, until now radiation-induced diffusion of defect clusters in bulk had not been demonstrated. This mechanism is consistent with our observation that decreasing the energy of the electron beam lowers the mobility of clusters because the electron-beam-induced atomic displacement cross section decreases as electron energy decreases [34]. In order to further demonstrate that ballistic collision can explain trends observed in our experimental data, we build a model of radiation-induced diffusion, which model relies on elastic electron-nucleus collision of the radiation beam with cluster atoms.

As shown in Ref. [8] and Fig. S2 in Ref. [22], a ground-state to ground-state migration of a cluster involves multiple steps. The multi-step nature of the migration process is further confirmed by our *ab initio* MD simulations of diffusion of a carbon tri-interstitial cluster [22]. In our radiation-induced diffusion model, we assume the cluster can perform a single step when it receives from energetic electrons an energy $E_T$ higher than a threshold energy $E_{th}$. This step might involve migrating one or more atoms at a time. The property $E_{th}$ here is different from the widely known threshold displacement energy [35], where the latter one is the minimum energy received by a lattice atom to be displaced into an interstitial site. Here $E_{th}$ is defined as the minimum energy received by cluster atoms to activate cluster diffusion and is assumed to be the same for all atoms in the cluster. The rate of steps $J$ induced by electron radiation can be estimated as $\Phi \times \sigma_{th} \times N_{atom}$. In this equation $\Phi$ is the electron flux, $\sigma_{th}$ is the electron-beam-induced displacement cross section for $E_T > E_{th}$ [22], and $N_{atom}$ is the number of atoms participating in the diffusion process of the cluster ($N_{atom}$ is included in the estimate of $J$ as we assume a step can be induced whenever any atom in the cluster receives $E_T > E_{th}$). Assuming that the number of steps to complete a migration between symmetry equivalent sites is $N_{step}$, the diffusion coefficient can be written as

$$D = \frac{a^2}{2d} \cdot \frac{J}{N_{step}} = \frac{a^2}{2d} \cdot \frac{\Phi \sigma_{th} N_{atom}}{N_{step}} = \frac{a^2}{2d} \cdot \frac{\Phi \sigma_{th}}{x}, \qquad (2)$$

where $d$ has the same meaning as in equation (1), $a$ is the migration distance between neighboring symmetry equivalent sites (~0.3 nm), and $x$ is the ratio of $N_{step}$ to $N_{atom}$. A reasonable range of values for the ratio $x$ is 0.5–2.0 [8,22,36]. At the sample surface the electron flux $\Phi_0$ is known, but dynamical diffraction of the electron wave in the crystal along the zone axis modulates the flux $\Phi$ as a function of sample thickness. Using a multislice simulation of electron beam in 4H-SiC we found that $\Phi$ can vary between 0.4 $\Phi_0$ and 1.4 $\Phi_0$ throughout the sample's depth [22]. As a cluster can exist at any depth, these minimum and maximum values of $\Phi$ are used to determine the lower and upper limits of the diffusion coefficient $D$.

In order to estimate $E_{th}$ we use the fact that in our experiments there is no observable displacement of clusters under 60 keV even at high electron flux. As shown earlier, the absence of observed diffusion implies the clusters have a diffusion coefficient smaller than $D = 3\times10^{-4}$ nm$^2$/s. Combining this value with $\Phi = 1.99\times10^7$ e$^-$/nm$^2$s (the maximum flux in the sample under 60 keV electron beam) and equation (2), one can estimate the lower limit of the threshold energy $E_{th}$ to be 10.7 eV. This value of $E_{th}$ is 2-3 times the typical migration barrier of small carbon interstitial clusters reported in literature (4.3–7.5 eV) [8]. This result is intuitive as the momentum transferred from the high-energy electron to a cluster atom will generally not align along the path of minimum migration energy for the excited step, yielding $E_{th}$ values significantly higher than typical migration energies.

Using the value of $E_{th}$ = 10.7 eV estimated from 60 keV experiments, we can now calculate the diffusion coefficient $D$ of clusters under 200 keV electron radiation. The results are shown as a function of the electron flux $\Phi$ in Fig. 3(b). For each value of $x$ the straight lines represent the maximum and minimum values of $D$ determined assuming $\Phi = 1.4\Phi_0$ and $\Phi = 0.4\Phi_0$, respectively. The diffusion coefficients predicted by the model are on the same order of magnitude as the ones measured experimentally, and a quantitative agreement can be reached for reasonable and physically-justifiable parameters of the model. Furthermore, the model predicts a linear increase of diffusivity with electron flux, similar to that seen in experimental data in Fig. 3(b), although within the error bars of the experimental data it is not obvious whether the trend is also linear. Finally, our model predicts that reducing the beam energy will lower $D$, since the electron-beam-induced atomic displacement cross-section $\sigma_{th}$ decreases monotonically with decreasing energy [34] and since $D$ is proportional to $\sigma_{th}$ (equation (2)). This prediction is again consistent with our experimental observation that diffusion coefficient for 60 keV electrons is significantly lower than for the case of 200 keV electrons. The results of the model strongly support our hypothesis that ballistic collision is the mechanism controlling experimentally observed radiation-induced mobility of self-interstitial clusters in SiC. A more quantitative analysis of diffusion coefficients would require performing experiments under other electron irradiation conditions, which are challenging to carry out in TEM. Nevertheless our experiments provide sufficient evidence for the ballistic collision mechanism proposed for the observed diffusion of defect clusters.

The combined results of STEM, *ab initio* MD simulations and electron beam-cluster collision model demonstrate that small interstitial clusters in SiC can become mobile under irradiation – contrary to what had been previously assumed. Based on the results of our 200 keV electron radiation experiments, the radiation-induced diffusion distance of these clusters is on the order of nanometers over hundreds of seconds. The distance can be larger if the materials are under radiation for longer times such as in the service of nuclear reactors. Such an enhanced mobility of clusters can potentially affect local defect dynamics including coalescence of neighboring clusters, recombination between clusters and vacancies as well as annealing of clusters at nearby sinks. Such processes can play a role in defect evolution in irradiated or ion-implanted materials because they impact the rate of damage accumulation and thereby also multiple properties of these materials relevant for their applications. It is also possible that one could use electron beam to anneal out undesirable interstitial clusters, which otherwise are known to persist even at high annealing temperatures [37]. While our measurements were carried out on SiC (a materials in itself important for many technological applications), we expect that similar behavior can be observed in other covalent or ionic-covalent materials if high resolution imagining techniques are used to monitor defect kinetics.

## Acknowledgment

200 keV STEM images were acquired under DOE NEUP grant # DE-NE0008418. Acquisition of 60 keV STEM images, all data analysis and simulations were supported by the US Department of Energy Basic Energy Sciences for funding this research (Fund number DE-FG02-08ER46493). The authors acknowledge A. B. Yankovich for help in setting up multi-slice calculations shown in Ref. [22].

**Figures**

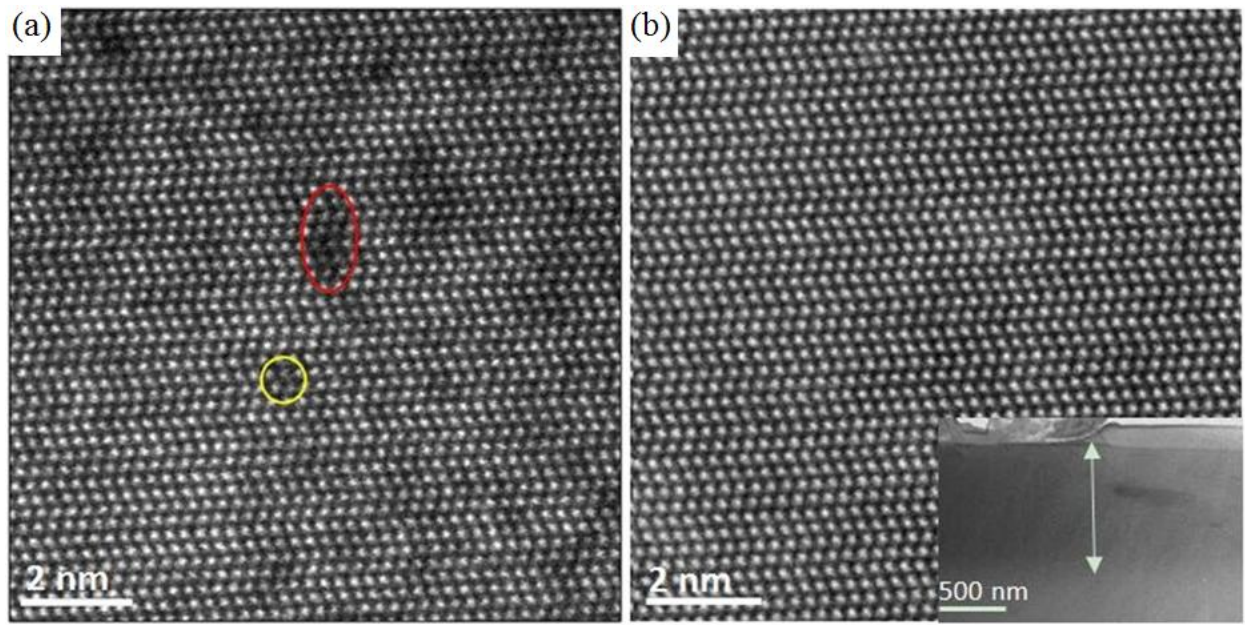

FIG. 1 (color online). (a), LAADF STEM image from the ion damaged region acquired with semi-convergence angle 17.5 mrad and collection angle 17.5 - 34 mrad shows defects induced strain contrast. Two example clusters have been circled, (b) HAADF STEM image of the same area as in (a) acquired with a beam semi-convergence angle 24.5 mrad and collection angle 54 - 270 mrad. The inset is a TEM bright field image with the viewing direction along $<11\bar{2}0>$. The arrow shows the ion damaged surface region. Both high-resolution images have been convolved with a 0.5 Å standard deviation Gaussian filter to reduce noise.

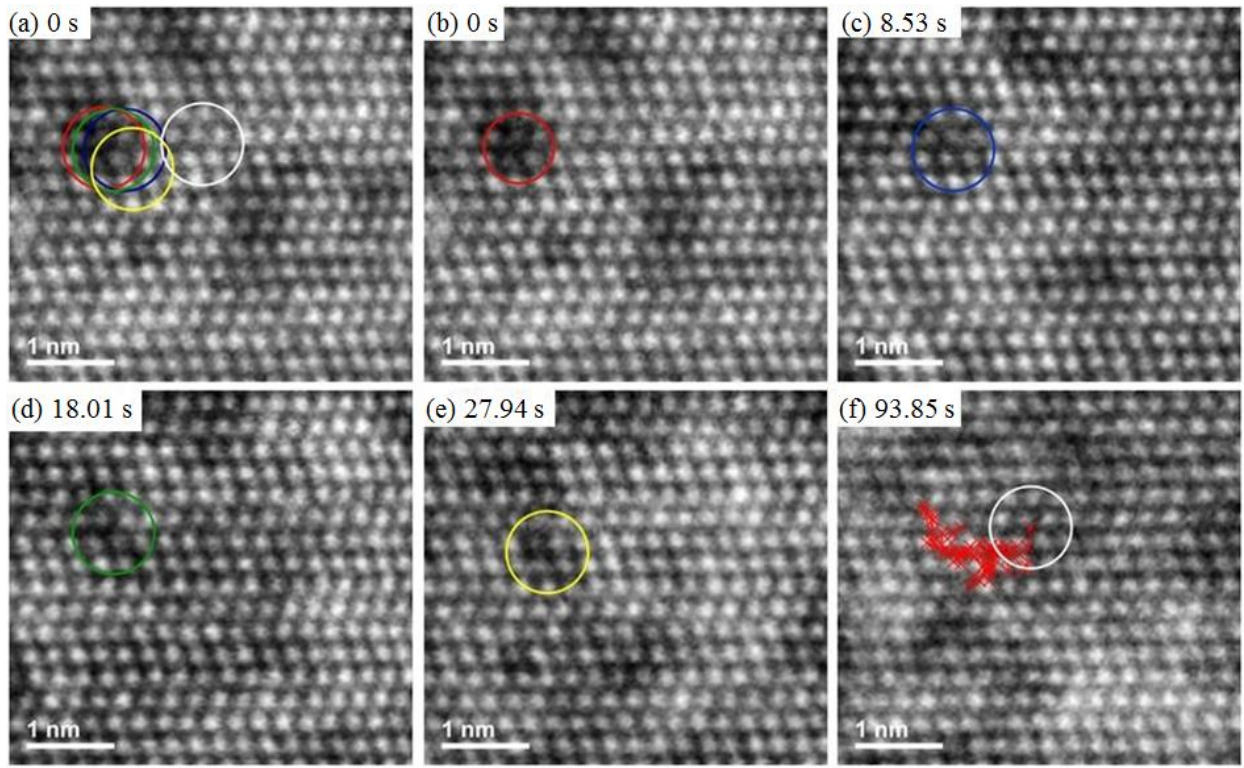

FIG. 2 (color online). Defect trajectory in an aligned STEM image series under $4.05\times10^6$ e$^-$/nm$^2$s 200 keV electron radiation. (a) The first image in the trajectory with the defect positions in the later images superimposed, (b)-(e) Snapshots of the defect position as a function of time, (f) The defect position at the end of the trajectory (white circle) and the entire trajectory (red symbols). Squared displacements were calculated from trajectories like (f). The entire trajectory is available as video S1 in Ref. [22]. Images were convolved with a 0.5 Å standard deviation Gaussian filter to reduce noise.

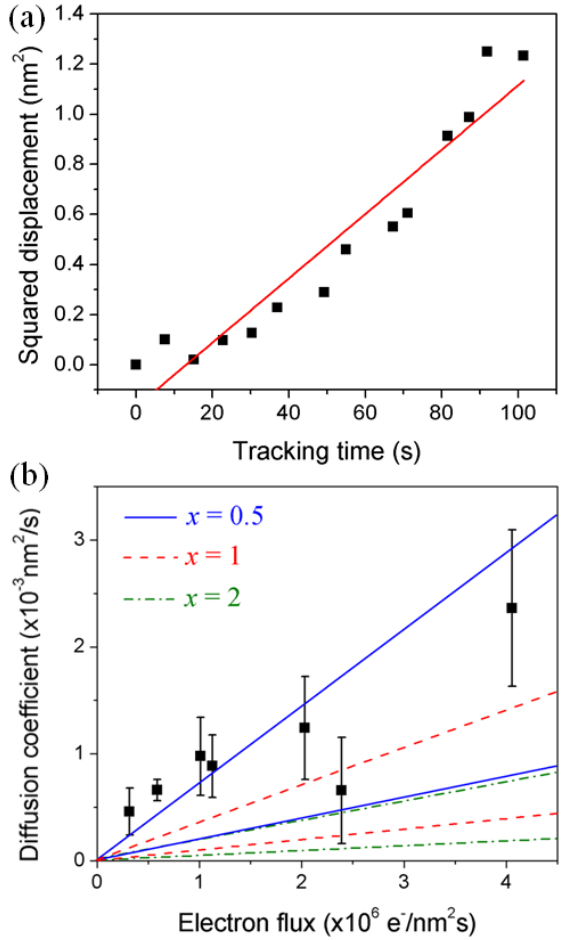

FIG. 3 (color online). (a) Mean square displacement (symbols) of the defect circled in Fig. 2 and a linear fit to the data, (b) Diffusion coefficients of mobile defect clusters measured at different fluxes under 200 keV electron beam. Each point represents the average over 3-5 mobile defects and the error bar is calculated as a standard deviation. Solid (blue), dashed (red), and dashed-dotted (green) lines indicate the upper and lower limits of diffusion coefficient predicted by the model using different values of $x$, where $x = N_{step}/N_{atom}$ as explained in text.